\RequirePackage{rotating}
\documentclass[useAMS,usenatbib]{mn2e}

\usepackage{amssymb,amsmath}
\usepackage{graphicx}
\usepackage{url}
\usepackage{ulem}
\usepackage{longtable}
\usepackage{multicol}
\usepackage{rotating,caption}

\newcommand{\be}{\begin{equation}}
\newcommand{\ee}{\end{equation}}
\newcommand{\ben}{\begin{equation*}}
\newcommand{\een}{\end{equation*}}

\newcommand{\splus}{\mbox{\raisebox{.4\height}{\scalebox{.4}{+}}}}
\newcommand{\sminus}{\mbox{\raisebox{.4\height}{\scalebox{.4}{-}}}}

\usepackage{color}

\title[Primordial shocks]
{State-to-state vibrational kinetics of H$_2$ and H$_2^+$ in a post-shock cooling gas with primordial composition}

\author[C.~M.~Coppola, G.~Mizzi, D.~ Bruno, F.~Esposito, D.~Galli, F.~Palla, S.~Longo]
{C.~M. Coppola$^{1,2,3}$
\thanks{e-mail: carla.coppola@chimica.uniba.it}, G.~Mizzi$^{4,5}$, D.~Bruno$^6$, F. ~Esposito$^6$, D.~Galli$^2$, F.~Palla$^2$, S. Longo$^{1,2,6}$ \\
$^1$Universit\`{a} degli Studi di Bari, Dipartimento di Chimica, Via Orabona 4, I-70126, Bari, Italy\\
$^2$INAF-Osservatorio Astrofisico di Arcetri, Largo E.~Fermi 5, I-50125 Firenze, Italy\\
$^3$Department of Physics and Astronomy, University College London, Gower Street, London WC1E 6BT\\
$^4$Centre for Complexity Science, Zeeman Building, University of Warwick, Coventry CV4 7AL, UK
$^5$Universit\`{a} degli Studi di Bari, Dipartimento di Fisica, Via Orabona 4, I-70126, Bari, Italy\\
$^6$IMIP-CNR, Section of Bari, via Amendola 122/D, I-70126 Bari, Italy}

\begin{document}

\date{}

\maketitle

\begin{abstract}
The radiative cooling of shocked gas with primordial chemical composition
is an important process relevant to the formation of the first
stars and structures, as well as taking place also in high velocity cloud
collisions and supernovae explosions.  Among the different processes
that need to be considered, the formation kinetics and cooling of molecular hydrogen
are of prime interest, since they provide the only way
to lower the gas temperature to values well below $\sim$10$^4$~K.
In previous works, the internal energy level structure of  H$_2$
and its cation has been treated in the approximation of 
rovibrational ground state at low densities, or trying to describe the dynamics
using some arbitrary $v>0$ H$_2$ level that is considered representative of the excited 
vibrational manifold. 
In this study, we compute the vibrationally resolved
kinetics for the time-dependent chemical and thermal evolution of 
the post-shock gas in a medium of primordial composition. 
The calculated non-equilibrium distributions
are used to evaluate effects on the cooling function
of the gas and on the cooling time. Finally, we discuss
the dependence of the results to different initial values of the shock velocity
and redshift.  
\end{abstract}


\section{Introduction}

Shock waves are commonly found in astrophysical processes involving
first structure formation, either galaxies or during the gravitational
collapse of the first stars. It is then interesting to investigate
the detailed microphysics for a gas of primordial composition. In
fact, in a shocked gas the cooling sensitively depends on
the radiative transitions of the chemical components and their
rapidly changing abundances. Atomic hydrogen and helium are very
inefficient at radiatively cooling below $\sim10^4$ K because in a
gas in collisional equilibrium they are usually completely recombined.
The lack of electrons and the cut-off in the functional dependence
of the Ly$\alpha$ excitation rate on temperature are responsible
for the steep drop of the cooling function in this temperature
regime. However, after the gas is shocked at temperatures far above
$10^4$ K, cooling proceeds faster than recombination, leaving
an enhanced ionisation fraction that promotes the formation of
molecular hydrogen through the intermediaries H$^-$ and
H$_2^+$. Furthermore, the presence of H$_2$ makes possible
further radiative cooling down to $\sim10^2$ K by ro-vibrational
line excitation, as discussed by e.g. \cite{shapiro_hydrogen_1987} and
\cite{machida_low-mass_2005}.

In this paper we consider the case of a strong, stationary and
completely ionising shock, as expected from
a supernova explosion or during the virialisation of a primordial density
perturbation (e.g. \cite{takizawa1998,birnboim2010}). The novelty of our approach is that the
network of vibrationally resolved chemical processes involving H$_2$
and H$_2^+$ in the post-shock region is included in the kinetics,
following the method described by \cite{c11}. Prior to this study,
the problem has been studied under various assumptions
concerning the internal level distribution of H$_2$; for example, \cite{shapiro_hydrogen_1987}
tried to model vibrational excitation inserting a representative excited level 
(more specifically, $v=6$), while 
\cite{johnson_cooling_2006} used the ground state approximation. 
This hypothesis was justified by the consideration that the
collision time between atomic and molecular hydrogen is far longer
than the decay time of the vibrational levels so that the excited
levels are radiatively depopulated. However, it has been
shown by \cite{c11} that a non-equilibrium distribution can arise because of the
specific features of the chemical formation pathway. For example,
during the recombination process, the formation of H$_2$ proceeds
via the highest excited levels. The resulting non-equilibrium
distribution function can then be used as the actual level population
at each time step to yield a more realistic description of the
cooling of the gas.

The paper is organised as follows: in
Section~\ref{methods} the equations and approximations used are
described, both for the dynamical and chemical model together with
the adopted cooling and heating functions. In Section~\ref{results}
the results obtained using different initial conditions are presented
and the calculated non-equilibrium distributions are described and
explained in the light of the detailed chemical processes.

\section{Formulation of the problem}
\label{methods}

\subsection{Jump conditions}

In our study, we consider shocks in a gas of primordial composition.
Thus, the gas mixture of the pre-shocked phase consists of H, He
and D nuclei with an abundance defined by the current
cosmological models.  The Rankine-Hugoniot jump conditions 
are derived assuming the
conservation of mass, momentum and energy between the pre- and
post-shock regions. Hereafter, each quantity related to the pre-
and post-shock is denoted with the subscript ``1'' and ``2'',
respectively. We assume that in the pre-shock phase the gas is
composed only by H, He and D atoms, with number density 
\be
n_1=n_{1,{\rm H}}+n_{1,{\rm D}}+n_{1,{\rm He}},
\ee
and that the temperature is low
enough that internal atomic excitation can be neglected. These
initial conditions are appropriate for most of the redshift interval
from the recombination to the epoch of formation of the first structures. 
In the following, the ratios between the initial fractional abundance
of H, He and D will be expressed by the symbols $x_{\rm H}$, $x_{\rm
He}$ and $x_{\rm D}$, respectively. The numerical values adopted are: 
$x_{\rm H}=0.924$, $x_{\rm He}=0.07599$, $[D/H]=2.6\times10^{-5}$. 
They have been obtained in the $\Lambda$CDM
cosmological model with implemented cosmological parameters according to \cite{planck_collaboration_2015}.

Behind the shock the gas density can be written as:
\be
n_2=n_{2,{\rm H}}+n_{2,{{\rm He}}}+n_{2,{\rm D}}+n_{2,{\rm H}^{\splus}}+n_{2,{{\rm He}^{\splus\splus}}}+n_{2,{\rm He}^{\splus}}+n_{2,{\rm D}^{\splus}}+n_{2,{\rm e}^{\sminus}}.
\ee
Defining the ionisation degrees of H, He, He$^+$ and D as $\alpha_1$, $\alpha_2$, $\alpha_3$ and $\alpha_4$, 
the corresponding Saha equations can be written as
\begin{equation}
n_{\rm e} \frac{\alpha_1}{1-\alpha_1}= \frac{2 g(\mathrm{H}^+)}{g(\mathrm{H})}n_0\cdot e^{-\frac{I_H}{k_BT_2}},
\label{saha1}
\end{equation}
\begin{equation}
n_{\rm e} \frac{\alpha_2(1-\alpha_3)}{1-\alpha_2}= \frac{2 g(\mathrm{He}^+)}{g(\mathrm{He})} n_0\cdot e^{-\frac{I_{He}}{k_BT_2}},
\label{saha2}
\end{equation}
\begin{equation}
n_{\rm e} \frac{\alpha_3}{1-\alpha_3}= \frac{2 g(\mathrm{He}^{++})}{g(\mathrm{He}^+)} n_0\cdot e^{-\frac{I_{He\splus}}{k_BT_2}},
\label{saha3}
\end{equation}
\begin{equation}
n_{\rm e} \frac{\alpha_4}{1-\alpha_4}= \frac{2 g(\mathrm{D}^+)}{g(\mathrm{D})} n_0\cdot e^{-\frac{I_{D}}{k_BT_2}},
\label{saha4}
\end{equation}
where 
\be
n_0\equiv \left(\frac{2\pi m_{\rm e} k_{\rm B} T_2}{h^2}\right)^{3/2}.
\ee
Here, $g$ represents the degeneracy of the nucleus and $n_{\rm e}$
is the total electron density deriving from all the ionisation
processes. These equations must be solved along with mass, momentum
and energy conservation
\be
\rho_1 v_1 =\rho_2 v_2,
\ee
\be
p_1+\rho_1 v_1^2 =p_2+ \rho_2 v_2^2,
\ee
\be
h_1+\frac{v_1^2}{2}=h_2+\frac{v_2^2}{2}.
\ee
Here $v$ is the gas velocity, $\rho_1$ and $\rho_2$ are the mass densities before and after the shock,
\be
\rho_1=n_{1,{\rm H}}m_{\rm H}+n_{1,{\rm He}}m_{\rm He}+n_{1,{\rm D}}m_{\rm D}\equiv n_1 m_1,
\ee
with
\be
m_1=x_{\rm H} m_{\rm H}+x_{\rm D} m_{\rm D}+x_{\rm He} m_{\rm He},
\ee
and
\be
\begin{split}
\rho_2=&n_{2,{\rm H}^{\splus}}m_{{\rm H}^{\splus}}+n_{2,{\rm He}^{\splus}}m_{{\rm He}^{\splus}}+n_{2,{{\rm He}^{\splus\splus}}}m_{{\rm He}^{\splus\splus}}+\\
       &+n_{2,{\rm D}^{\splus}}m_{{\rm D}^{\splus}}+n_{2,{\rm e}^{\sminus}}m_{{\rm e}^{\sminus}},
\end{split}
\ee
and $p$ is the gas pressure given by the ideal gas law
\be
p_i=n_ik_{\rm B}T_i ,\qquad i=1,2.
\ee
Finally, the enthalpy per unit mass $h$ is defined as
\be
h_1=\frac{5}{2}\left(\frac{n_1k_BT_1}{\rho_1}\right),
\ee
\be
\begin{split}
h_2&=\frac{5}{2}\left(\frac{n_2k_BT_2}{\rho_2}\right)\\
   &+\frac{n_{2,{\rm H}^{\splus}}I_{\rm H}+n_{2,{\rm D}^{\splus}}I_{\rm D}+n_{2,{\rm He}^{\splus}}I_{\rm He}+n_{2,{\rm He}^{\splus\splus}}I_{\rm He^{\splus}}}{\rho_2},
\end{split}
\ee
where $I_{\rm H}$, $I_{\rm D}$, $I_{\rm He}$ and $I_{\rm He^{\splus}}$ are the ionisation potentials
of H, D, He and He$^{\splus}$.

Defining $x\equiv \rho_2/\rho_1$ as the compression ratio and
combining the equations for the conservation of momentum and energy,
the relationship between the pre- and post-shock densities becomes
\be
4v_1^2x^{-2}-5\left(v_1^2+\frac{k_{\rm B}T_1}{m_1}\right)x^{-1}+v_1^2+\frac{5k_{\rm B}T_1}{m_1}-2I=0
\ee
where the term $I$ takes into account the energy used for the ionisation of H, D and He in the shock front,
\be
I=\frac{x_{2\mathrm{H}^+}I_H}{m_2}+\frac{x_{2\mathrm{D}^+}I_D}{m_2}+\frac{x_{2\mathrm{He}^+}I_{He}}{m_2}+\frac{x_{2\mathrm{He}^{++}}I_{He^+}}{m_2}.
\ee
The pre-shock density of the gas is calculated at a
redshift $z$, assuming a
uniform intergalactic medium with density $n_{\rm H}\approx 0.19 (1+z)^3$~m$^{-3}$.
As an illustration, at $z=10$, $n_{\mathrm{H}}=250$ m$^{-3}$.  
Neglecting the trivial solution (for which
the density after the shock is equal to the density before the
front), we obtain for the compression ratio the value $x=5$,
close to the asymptotic value (for strong shocks $x=4$).

\subsection{Chemical network}

\begin{figure}
\begin{center}$
\begin{array}{l}
\includegraphics[width=6cm, angle=270]{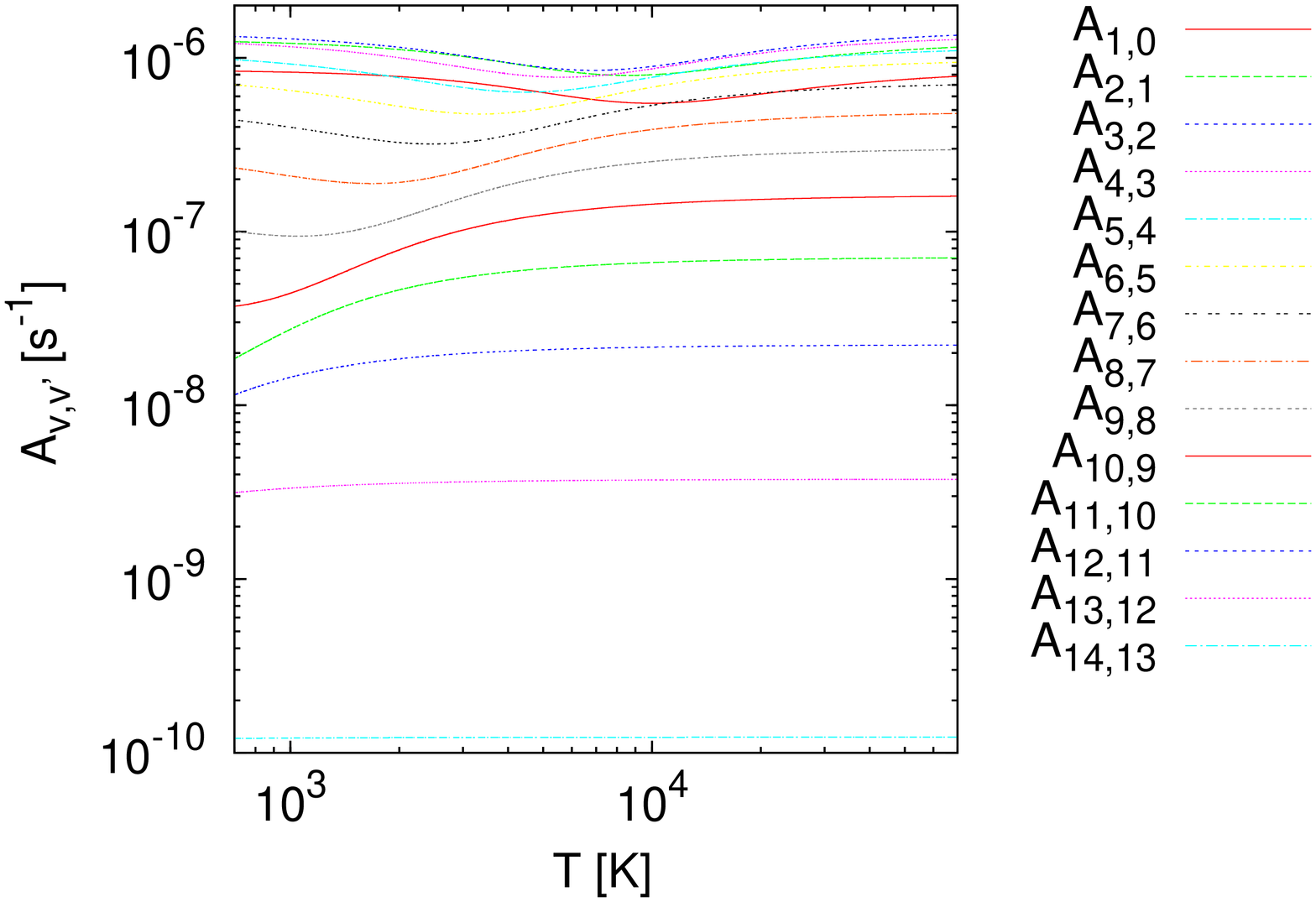}\\
\includegraphics[width=6cm, angle=270]{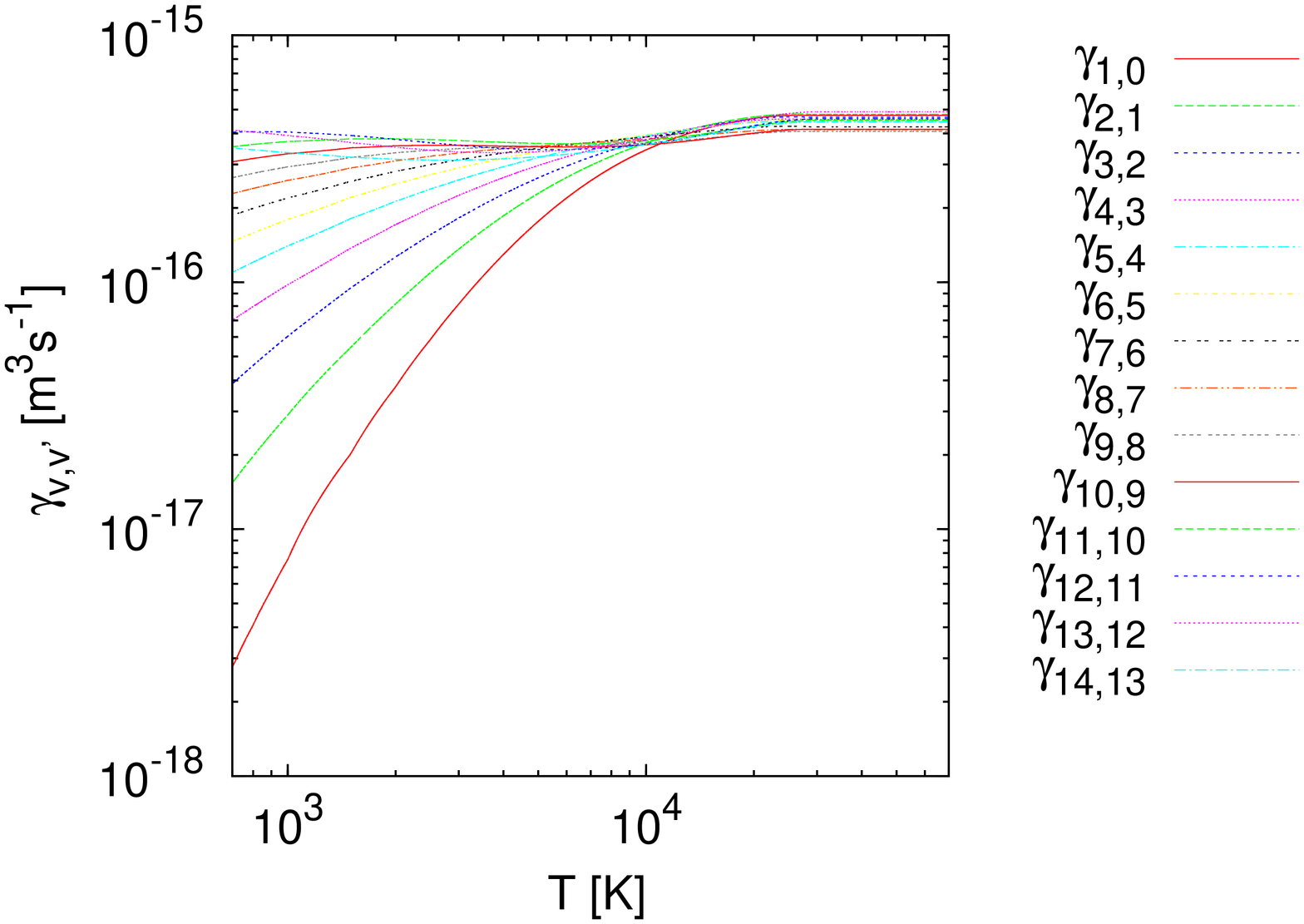}\\
\end{array}$
\end{center}
\caption{Evolution of state-resolved thermally-averaged Einstein coefficients $A_{v,v'}$ and 
collisional coefficients $\gamma_{v,v'}$ as a function of $T$;
among the full set of transitions included in the kinetic model, 
only the coefficients for the transitions with $\Delta v=1$ are shown here as an example. 
}
\label{avv}
\end{figure}

We consider the following species: e$^-$, H, H$^+$, H$^-$, H$_2$,
H$_2^+$, D, D$^+$, D$^-$, He, He$^+$, He$^{++}$, HD, HD$^+$, HeH$^+$,
H$_3^+$. In the first phase of the shock the helium chemistry plays
an important role and the kinetics of formation and destruction of
HeH$^+$ has been introduced to follow the recombination of helium
cations. The rate coefficients for processes not involving molecules
are listed in Table~\ref{tab:atomic}, while those for molecular
processes (including formation, destruction and redistribution among
the vibrational manifolds) have been taken from \cite{c11}
(see Table~\ref{tab:molecular}). The ground electronic states of
H$_2$ and H$_2^+$ support 15 and 19 bound vibrational levels including the
vibrational ground state, respectively.

The major formation pathway for $\mathrm{H}_2^+$ is the radiative
association  \citep{mihajlov_2007, mihajlov_2014}.
\be
\mathrm{H}^++\mathrm{H}\rightarrow\mathrm{H}_2^++h\nu,
\ee
while destruction is mainly achieved through dissociative recombination \citep{takagi_2005, motapon_2008, motapon_2014}.
\be
\mathrm{H}_2^++\mathrm{e}^- \rightarrow\mathrm{H}+\mathrm{H},
\ee
photodissociation \citep{mihajlov_2007, mihajlov_2014}.
\be
\mathrm{H}_2^++h\nu \rightarrow\mathrm{H}+\mathrm{H}^+,
\ee
and charge transfer \citep{krstic_2002,krstic_2002b}.
\be
\mathrm{H}_2^++\mathrm{H} \rightarrow\mathrm{H}_2+\mathrm{H}^+,
\ee
The main formation channels for H$_2$ include associative
detachment \citep{cizek_1998,kreckel_2010}
\be
\mathrm{H}+\mathrm{H}^- \rightarrow\mathrm{H}_2+\mathrm{e}^-
\ee
and charge transfer \citep{krstic_2002,krstic_2002b}
\be
\mathrm{H}_2^++\mathrm{H} \rightarrow\mathrm{H}_2+\mathrm{H}^+
\ee

All the vibrationally resolved processes introduced are listed in
Table~\ref{tab:molecular} following the model by \cite{c11} where
an analysis of each process can be found. It should be noted that
both collisional excitations and de-excitations for H impact 
and spontaneous radiative transitions have
been included in the kinetic model as well as in the cooling and
heating functions. For some specific channel where no
state-to-state data have been found, the nascent distributions
have been estimated according to the exothermicity
of the process and on available information. For H$_2$ and H$_2^+$
recombination the work by \cite{h3+_h2+} has been considered to
justify the formation mainly in higher vibrational levels.
It is important to mention that for the chemical processes involving
H$_3^+$ the vibrational levels of H$_2$ and H$_2^+$ have been taken
equal to the highest levels since the recombination most likely
proceeds in this way (e.g. \cite{h3+_h2+}); however, accurate state-to-state calculations
would be necessary for the proper description of the chemical kinetic effects due to this channel.

To study the evolution of the chemical composition in the post-shock gas, the
system of ordinary differential equations must be solved:
\be
\begin{split}
\frac{dn_j}{dt}=&-n_j\sum_{j'}(R_{jj'}+P_{jj'}+n_{\rm H}\gamma_{jj'})+\\
&+\sum_{j'}R_{jj'}n_{j'}+\sum_{j'}\sum_{j''}\mathbf{C}_j^{j'j''}n_{j'}n_{j''}
\end{split}
\label{odesys}
\ee
where the symbols have the following meaning:
\begin{itemize}
\item $n_j$ are the densities of the various species. In this
notation, different vibrational levels of $\mathrm{H}_2$ and
$\mathrm{H}_2^+$ have a different index $j$;
\item $R_{jj'}$ are the spontaneous and stimulated excitation and de-excitation rates,
that can be expressed in terms of the Einstein coefficients $A_{jj'}$ as 
\be
R_{jj'}=
\begin{cases}
A_{jj'}[1+\eta_{jj'}(T_{\rm r})] & \text{if } j>j',\\
(g_{j'}/g_j) A_{jj'}\eta_{jj'}(T_{\rm r}) & \text{if } j<j'
\end{cases}
\ee
where the statistical weight of the molecular vibrational levels
is $g_j=1$ since there is no vibrational degeneracy
and  $\eta_{jj'}(T_{\rm
r})=[\exp({h\nu_{jj'}/k_{\rm B}T_{\rm r}})-1]^{-1}$, where $T_{\rm r}=2.73 (1+z)$~K is 
the temperature of the cosmic background (CMB);
\item $P_{jj'}$ are the destruction rates of the $j^{th}$
species due to photons;
\item $\gamma_{jj'}$ are the excitation/de-excitation rate coefficients of the $j^{th}$
species for collision with H;
\item $\mathbf{C}_j^{j'j''}$ are the formation rates
of the $j^{th}$ species for collision between the $j'^{th}$ and
$j''^{th}$ species, photons included.
\end{itemize}

Since we do not resolve the molecular rotational energy levels, we have computed the Einstein coefficients 
for the vibrational transitions by averaging the fully resolved $A_{(v,J)\rightarrow (v^\prime, J^\prime)}$ 
computed by \cite{wolniewicz_1998} over a thermal distribution of rotational level populations. As a consequence,
the $A_{vv'}$ are functions of the temperature, like the collisional rate coefficients $\gamma_{vv'}(T)$.
For the latter, at high temperatures 
we have adopted the results obtained by \cite{esposito2009} using the quasiclassical trajectory method. It is well know that
at temperatures below $\sim 500$--600~K quantum-mechanical effects
should be taken into account. Thus, we have included in the kinetics 
the data by \cite{flower_1997} and \cite{flower_1998} (24 ortho- and 27 para- transitions) down
to temperatures of 100~K. All the data available for rovibrational transitions have 
been adopted averaging on the rotational distribution. At lower temperatures, data have been extrapolated using an Arrhenius-type
law. Fig.~\ref{avv} shows the Einstein coefficients $A_{vv'}(T)$ and the 
collisional rate coefficients $\gamma_{vv'}(T)$
as function of the temperature. 

The full system of equations has been numerically solved. The rate
coefficients for atomic processes are computed as function of
the temperature of the gas for collisional processes and of the
temperature of the CMB for processes involving photons. For molecular
processes, the rate coefficients are computed by interpolating
state-to-state calculated rate coefficients.  The various contributions
to the cooling function are computed from analytical expressions, that are reported
in Table \ref{tab:cooling} together with the corresponding references. 
The fractional abundance of electrons is obtained
by imposing charge neutrality.  Other contributions to the cooling
function due to atomic processes are listed in Table~\ref{tab:cooling}.


\subsection{Cooling and heating}
\label{sec:cooling_intro}

It is useful to summarise the conservation laws that hold between
the pre- and post-shock quantities. Behind the shock front conservation
of mass and momentum still applies
\be
\rho\frac{dv}{dt}=-v\frac{d\rho}{dt},
\label{mass_flux_der}
\ee
\be
\frac{dp}{dt}=v^2\frac{d\rho}{dt},
\label{momentum_flux_der}
\ee
where we have dropped the subscript 2 for clarity.
The energy equation can be written as an equation for the evolution of temperature:
\begin{equation}
\frac{dT}{dt}=(\gamma-1)\left(\frac{\Gamma-\Lambda}{nk_{\rm B}}+\frac{T}{n}\frac{dn}{dt}\right)
+\gamma\frac{T}{\mu}\frac{d\mu}{dt}.
\label{tempeq}
\end{equation}
where $\mu$ is the mean atomic weight.
We assume that $\gamma$ is constant and equal to the monoatomic value $\gamma=5/3$.
This is justified by the fact that the fractional abundance of
molecular hydrogen is always negligible and the gas can essentially
be regarded as monoatomic.

Eq.~(\ref{tempeq}) can be simplified using some approximations. 
The last term is usually negligible since it changes primarily because of recombination. 
In addition we can assume that the pressure is constant. Indeed, previous work 
\citep{shapiro_hydrogen_1987} has shown that the pressure varies 
only slightly, as confirmed by our complete calculations. With these assumptions,
the equation for the gas temperature can be simplified as
\be
\frac{dT}{dt}=\frac{\gamma-1}{\gamma}\left(\frac{\Gamma-\Lambda}{nk_{\rm B}}\right).
\label{temp_der}
\ee
If $T \gg 10^4$~K a shocked gas cools faster than it recombines.
Thus, recombination is not in equilibrium and there is a significant
ionisation fraction below $10^4$~K which allows H$^-$ and H$_2^+$
to form and eventually produce H$_2$ molecules.  Even though its
abundance is small, H$_2$ is the only possible cooling agent down
to $\sim100$~K by means of rovibrational line excitation; for the 
same reasons, at lower temperatures HD allows to cool even more 
efficiently \citep{shchekinov_formation_2006,johnson_cooling_2006}. In these
conditions the gas is completely neutral and H$_2$
excitation occurs only via collisions with H. 
Radiative cooling functions computed in the LTE approximation are available for some of the molecules
included in the present model, namely HD$^+$, HeH$^+$ and H$_3^+$
\citep{coppola_2011,miller_2010}. However, the critical densities
for most of the transitions of these molecules are much higher than
the density regime covered by our calculations and for this reason
they have not been included.

Several cooling mechanisms have been included in the model, both involving atoms/ions and molecules. 
Among the former, the following processes have been inserted:
\begin{enumerate}
\item radiative cooling from collisional excitation of H and He$^+$;
\item cooling by H and He collisional ionization;
\item bremsstrahlung;
\item Compton cooling;
\item cooling by recombination of H$^+$, He$^+$ and He$^{++}$.
\end{enumerate}
while for the latter, we have considered:
\begin{enumerate}
\item cooling for the collisions: H$_2$/H, H$_2$/H$_2$, H$_2$/He, H$_2$/H$^+$ and~H$_2$/e$^-$;
\item HD cooling.
\end{enumerate}

Concerning the contribution of HD cooling, we have included the biparametric fit provided
by \cite{lipovka_cooling_2005}. In order to extend the calculation to higher $z$ the cooling function for HD should be corrected:
in fact, the fit provided by \cite{lipovka_cooling_2005} has been calculated at $z=0$.

Two other processes have also been included in the thermal energy balance,
namely the heating due to the formation of H$_2$ via H$^-$ and the
cooling due to the electron attachment in the formation of the negative ion H$^-$.
For the former, the specific reaction rates $k_{ass\_det}^{v_i}$ for each vibrational
level have been used in the calculation:
\begin{equation}
\Gamma=\sum_{v_i} k_{ass\_det}^{v_i}n_Hn_{H^-}(E_{v_i}-E_0)
\label{gamma}
\end{equation}
The details about the implemented fits for each cooling/heating contribution and their 
references can be found in Table~\ref{tab:cooling}.

\begin{figure}
\includegraphics[width=6cm,angle=-90]{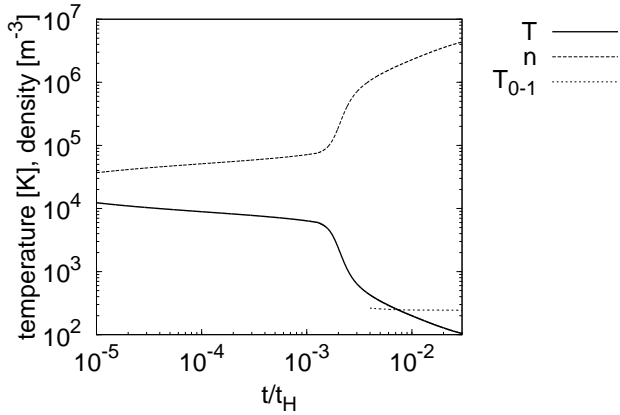}
\caption{Temperature and density profiles as a function of time; the value
of the excitation temperature of the lowest vibrational transition $T_{1-0}$ 
is shown for comparison (see the comments given in Sec.~\ref{results}).}
\label{fig:temp}
\end{figure}

The resulting trend for gas temperature and density is shown in Fig.~\ref{fig:temp}


\section{Results}
\label{results}

As a standard case we consider here the shock produced by a supernova
explosion with a shock velocity $v_s=50$ km~s$^{-1}$ and a pre-shock hydrogen
density corresponding to the baryon density at $z=20$. For $v_s \lesssim 50$ km~s$^{-1}$, 
 the post-shock radiation can be neglected \citep{shapiro_hydrogen_1987}. Moreover, together with the information given in Section \ref{methods},
the pre-shock fractional abundances have been introduced taking into 
account that the gas in the intergalactic medium before the formation of the 
first stars has a residual fractional ionization and a fraction of H$_2 \sim  10^{-6}$. 
Accordingly to the results shown in \cite{coppola_2011} we assumed this fraction to be in the ground
vibrational level. The pre-shock
temperature is given by the gas temperature at the value of $z$ at 
which the calculation is carried out, 
$T_1=T(z)$.
The post-shock ionization degrees have been calculated 
according to Saha's equations (Eqs.~\ref{saha1}-\ref{saha4}) coupled to 
mass, momentum and energy convervation equations. Including 
the terms corresponding 
to hydrogen, helium and deuterium ionisation in the energy conservation equation,
we obtain a post-shock temperature $\sim
6\times 10^5$~K. 

\begin{figure}
\includegraphics[width=8cm]{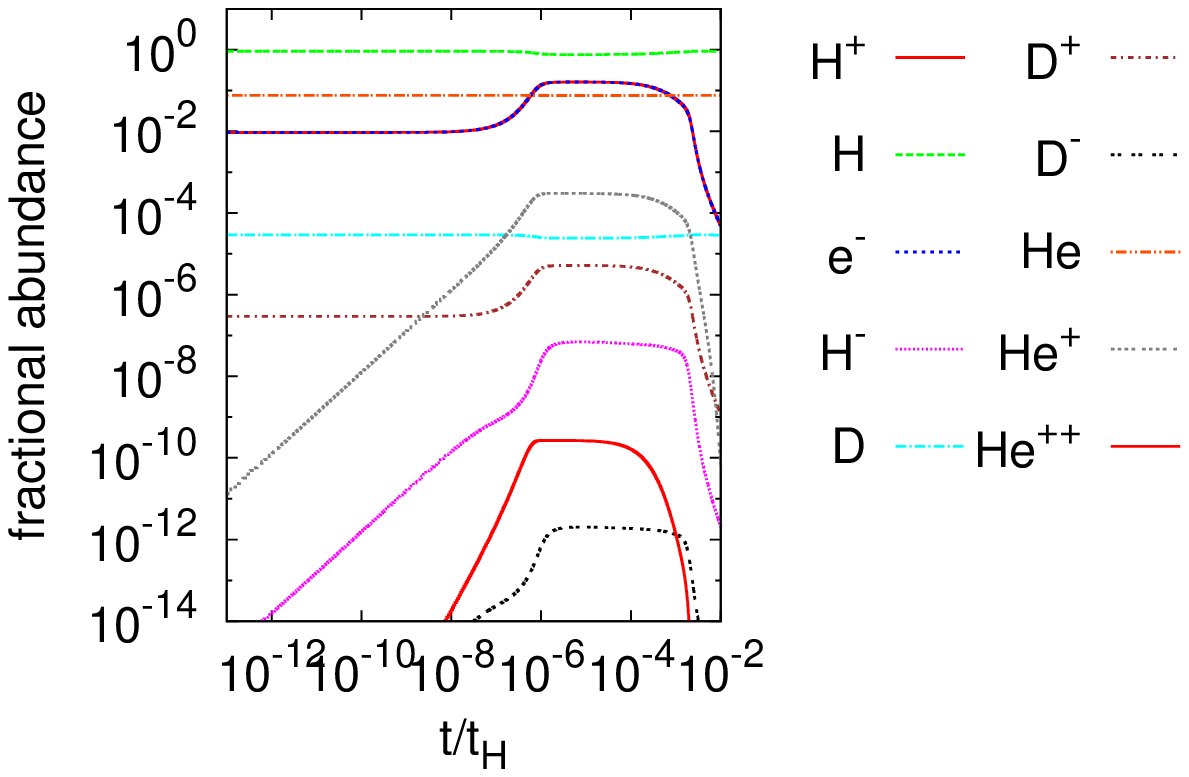} 
\includegraphics[width=8cm]{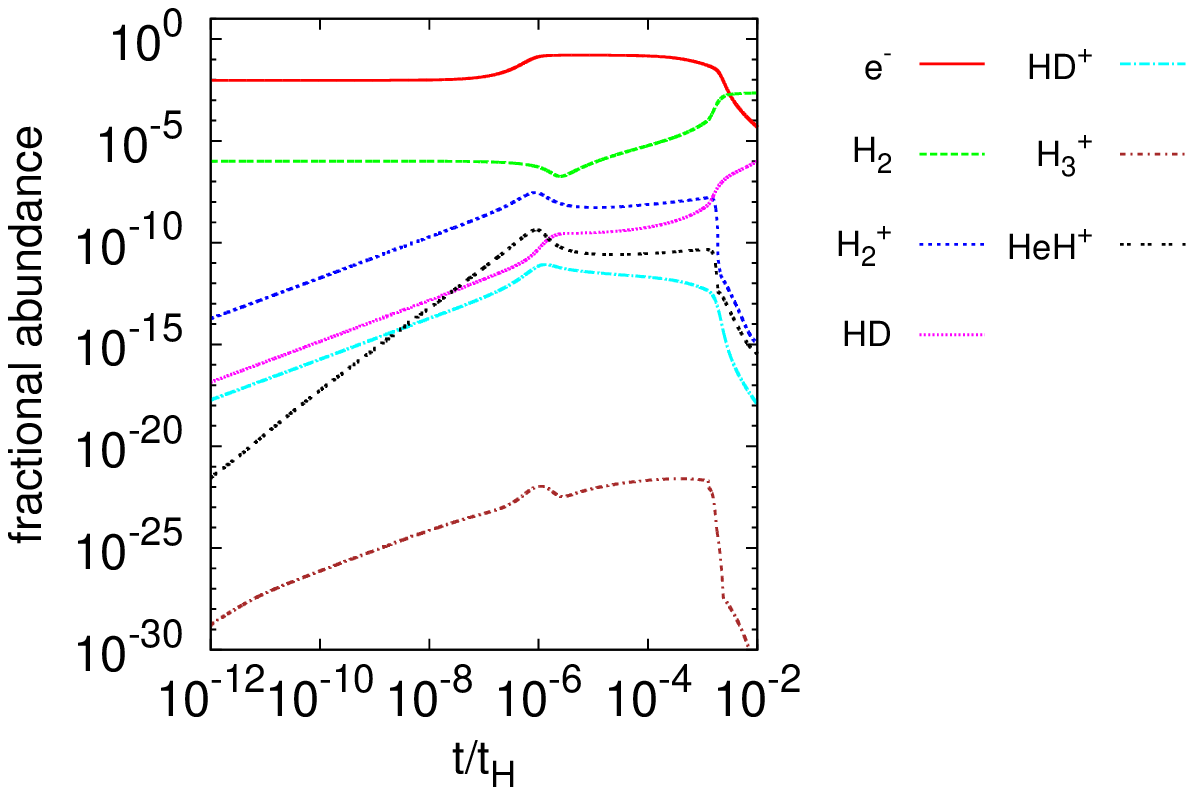} 
\caption{Fractional abundances of selected species. {\it Top panel}: atoms. 
{\it Bottom panel}: molecular species. For vibrationally resolved species only the abundance in the $v=0$ level 
is shown.}
\label{abundances_H}
\end{figure}

The chemical composition of the post-shock gas is shown in Fig.~\ref{abundances_H}
as function of time after the shock in units of the Hubble time, $t_{\rm H}\equiv H_0^{-1}=4.59\times 10^{17}$~s.
The latter is computed using the most recent value of the Hubble constant $H_0$ from \cite{planck_collaboration_2015}. Starting from an
initially ionised atomic gas, the formation of several molecules
and of the negatively charged hydrogen ion is observed to occur at
$t/t_{\rm H}\sim 0.003$ following the recombination epoch.

\begin{figure*}
\includegraphics[width=6.5cm,angle=-90]{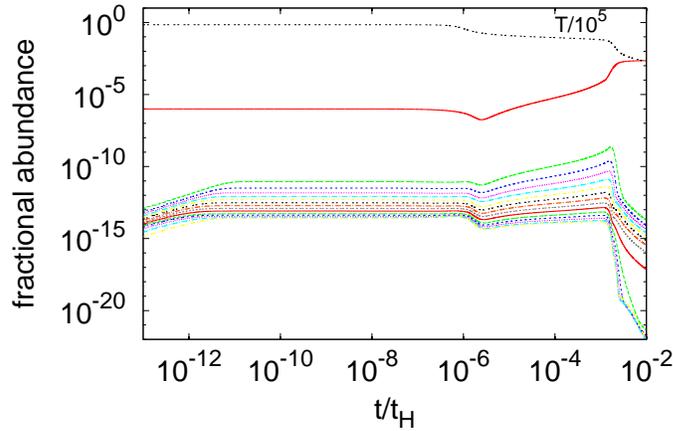}
\caption{Fractional abundance of H$_2$, from $v=14$ (bottom curve) to $v=0$ (top curve). 
The gas temperature is also shown as a reference.}
\label{abundpaper}
\end{figure*}

\begin{figure*}
\includegraphics[width=6.5cm,angle=-90]{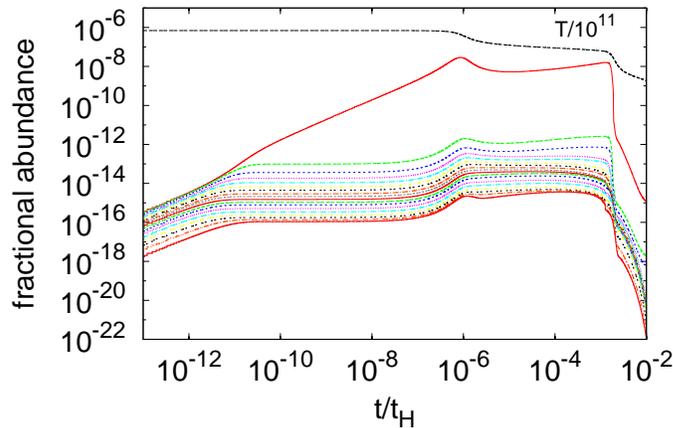}
\caption{Fractional abundance of H$_2^+$, from $v=18$ (bottom curve) to $v=0$ (top curve).
As in Fig.~\ref{abundpaper} the gas temperature is shown as a reference,
but on a different scale.}
\label{abundpaper2}
\end{figure*}

\begin{figure*}
\includegraphics[width=\columnwidth,angle=-90]{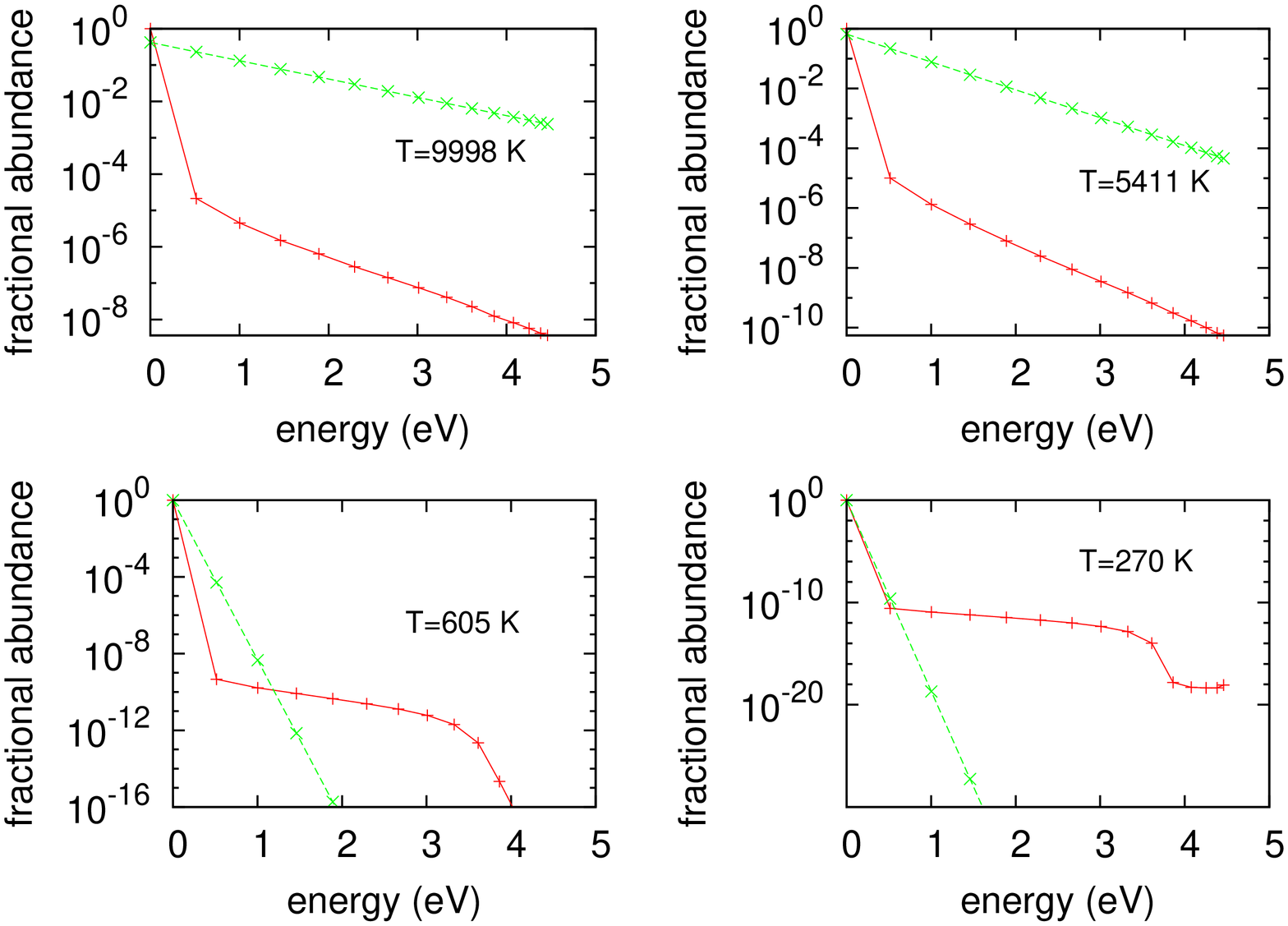}
\caption{Vibrational distribution function of $\mathrm{H}_2$ at different gas 
temperatures, corresponding to different times of the evolution (solid red curves). 
The equilibrium vibrational distribution function at the gas temperature
at the same time of the time evolution of the shock is
also shown (dotted green lines with crosses). At longer times from
the initial shock (i.e. at lower gas temperatures), non-equilibrium features
still characterise the shape of the distribution; in particular,
supra-thermal tail can be noticed due to recombination processes
occurring in the system.}
\label{fig:boltH2}
\end{figure*}

\begin{figure*}
\includegraphics[width=\columnwidth,angle=-90]{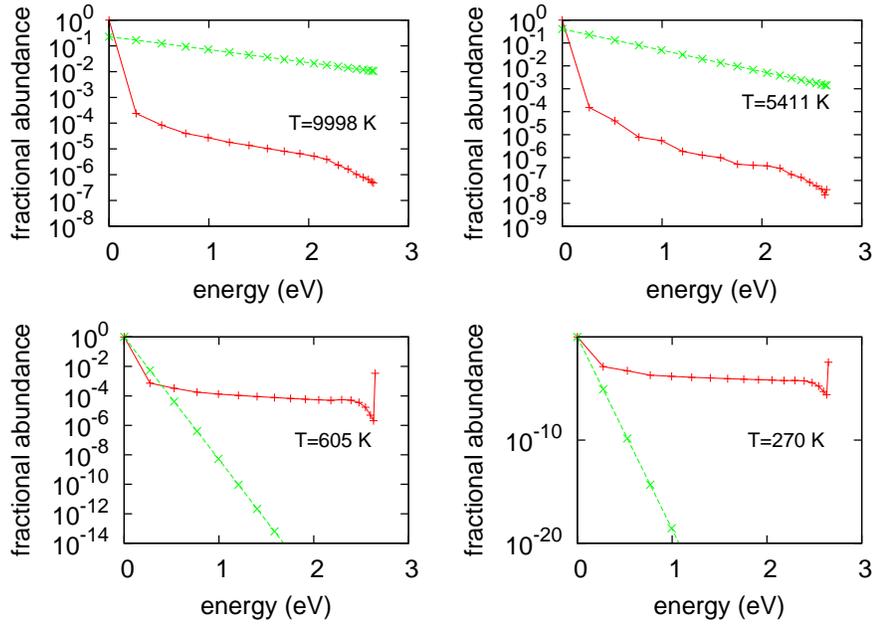}
\caption{As Fig.~\ref{fig:boltH2} but for H$_2^+$.}
\label{fig:boltH2p}
\end{figure*}

Fig.~\ref{abundpaper} and \ref{abundpaper2} show the evolution
of the populations of the vibrational manifolds of H$_2$
and H$_2^+$, while Figs. \ref{fig:boltH2} and
\ref{fig:boltH2p} show the non-equilibrium distributions of vibrational
levels of H$_2$ and H$_2^+$
compared to the Boltzmann distributions at the gas temperature
of the corresponding times in the evolution of the shock. As in the case of
the primordial Universe chemistry, it is possible to see two steps
in the evolution of the fractional abundance of H$_2$: the former is related to the formation through the charge transfer between
H$_2^+$ and H at $\sim 10^{-3}$ and $\sim 10^{-2}$ Hubble times, corresponding
to a gas temperature between $10^6$~K and 20,000~K. Indeed, a rise
in the H$_2^+$ fractional abundance can be seen even at slightly
earlier phases. The second step is due to the H$^-$ channel. Note
the drop of the fractional abundances of the vibrational levels
population for levels higher than $v=9$. This is due to the characteristic
endothermicity of the associative detachment process for highly
excited levels, as pointed out in the case of the early Universe
chemistry by \cite{c11}. The supra-thermal tails arise because of recombination
processes: starting from an initial phase of under-populated levels,
an increased production of higher excited levels can be noted. The
temperature associated with the first vibrational transition is
significantly smaller than the temperature associated with the tail
of the distribution. 

Concerning the H$_2^+$ vibrational level population it can be noticed
that the formation channel via HeH$^+$ 
is responsible for the bumps present at later times in the shock evolution ($t/t_H>4\times10^{-3}$).
In the present simulation, we assume that only the first 3 levels of the molecular hydrogen cation 
are formed via HeH$^+$ and that they are produced with the same specific rate.
A detailed state-to-state
calculation could in principle discriminate among the levels and provide a more accurate description for the chemical 
kinetics; for this, new data are needed.

\begin{figure}
\includegraphics[width=9cm]{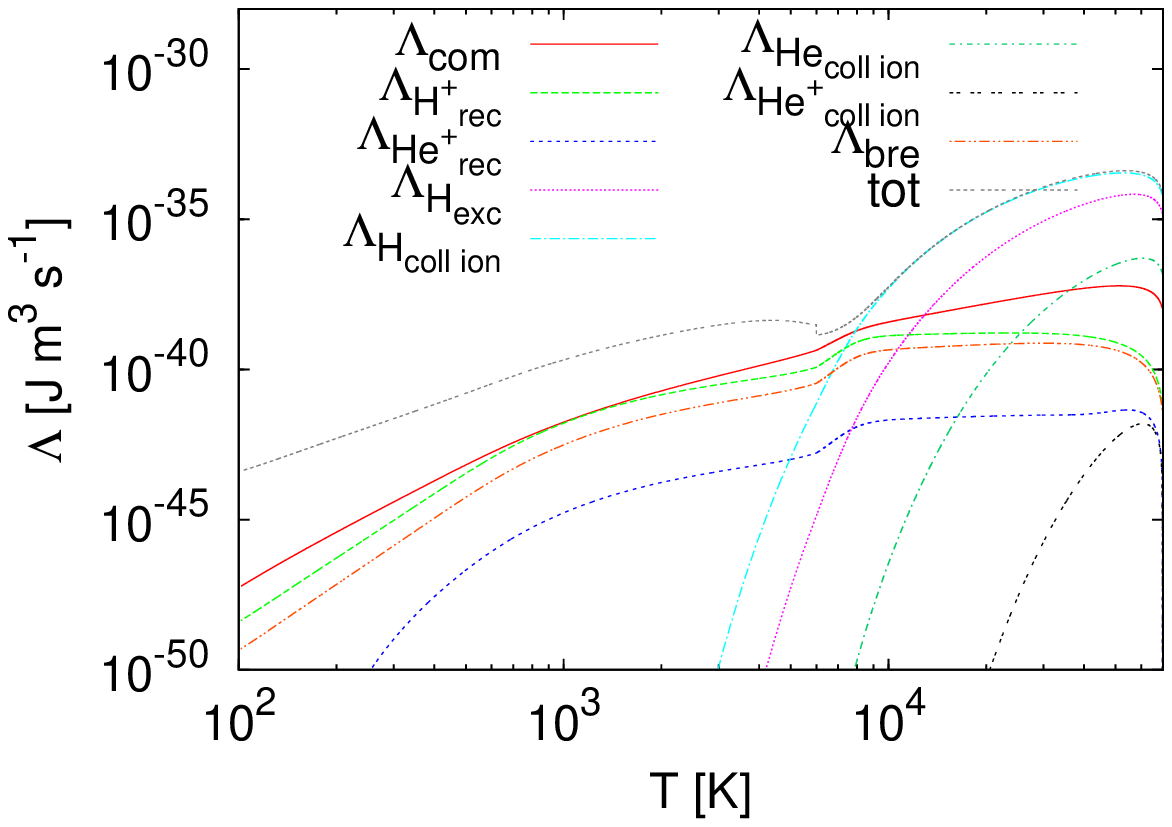}
\includegraphics[width=9cm]{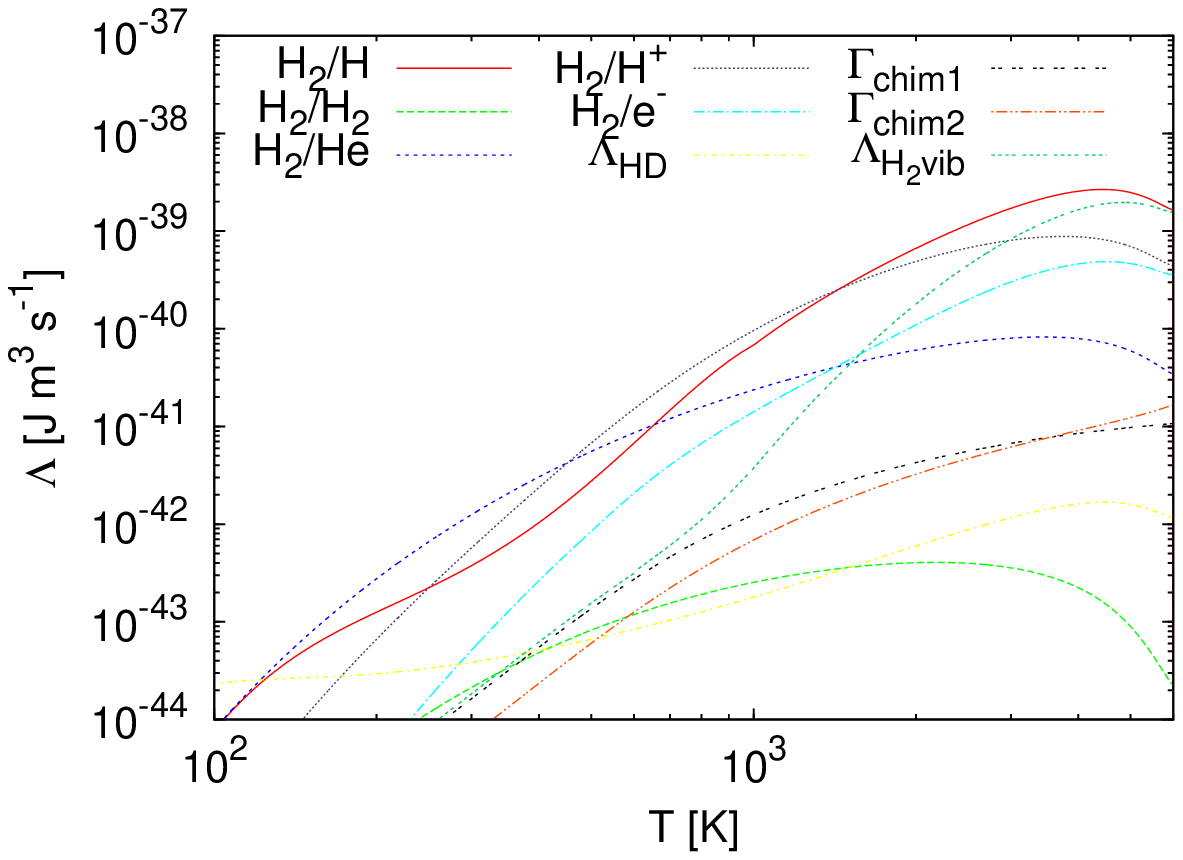}
\caption{Cooling functions for the post-shock gas. {\it Top panel}: atomic, ionic, 
free-electrons (Compton and Bremsstrahlung) and H$^+$/He$^+$/He$^{++}$ cooling functions.
{\it Bottom panel}: molecular contributions; namely, the H$_2$ cooling functions for
collisions with H, H$_2$, He \protect\citep{glover2008}, H$^+$ and e$^-$ 
\protect\citep{glover2015} (the last four contributions have been
included in the equation
for the evolution of the temperature); H$_2$ non-equilibrium vibrational cooling function
(see the text for the details) and HD cooling function by 
\protect\cite{lipovka_cooling_2005}).  The chemical contribution to the heating due
to the associative detachment of H$^-$ ($\Gamma_1$) and H$^-$ formation due to electron attachment
($\Gamma_2$) are also reported and included in the model.}
\label{coolingfunctions}
\end{figure}

\begin{figure}
\includegraphics[width=\columnwidth]{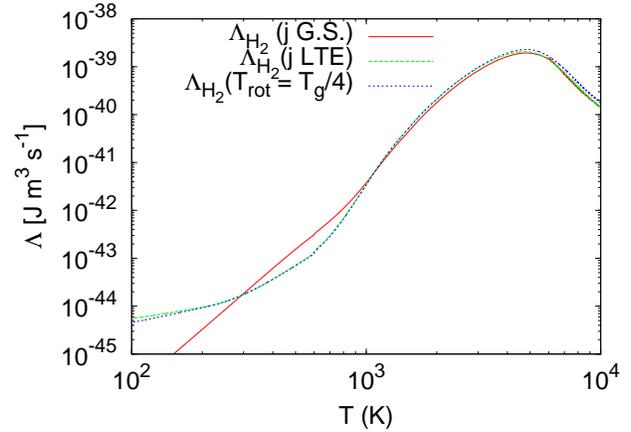}
\caption{H$_2$ cooling function: comparison among different approximations for the rotational levels of H$_2$. 
Three cases have been reported: (i) LTE at gas temperature (ii) LTE at a gas temperature 4 times smaller than the gas
temperature at each time (iii) ground state (G.S.) approximation.}
\label{figcomp}
\end{figure}

As a comparison with previous H$_2$ cooling calculations, 
the non-equilibrium cooling function and the fit provided by \cite{glover2008}
and \cite{glover2015} are compared in Fig.~\ref{coolingfunctions}. The non-equilibrium
cooling function has been calculated adopting the usual expressions \citep{tine_1998,omukai_2000}:
\be
\Lambda=\sum_{E_{v_i,j_i}>E_{v_f,j_f}} n_{v_i,j_i} A_{(v_i,j_i)\rightarrow (v_f,j_f)}(E_{v_i,j_i}-E_{v_f,j_f})
\ee
  where $n_{v_i,j_i}$ is the density of the initial roto-vibrational state,
 $A_{(v_i,j_i)\rightarrow (v_f,j_f)}$ is the spontaneous transition probability between an 
 initial ($v_i,j_i$) rotovibrational state to a final ($v_f,j_f$) one,
 $E_{v_i,j_i}$ and $E_{v_f,j_f}$ the energies of the initial and final state, respectively.
The cooling function calculated according to this procedure can't probe any rotational
non-equilibrium
effects. However, since a complete set of rotationally-resolved reaction rates 
for the most relevant 
processes involving H$_2$ is not available, we assumed that the populations of the rotational
levels follow
a Boltzmann distribution at the gas temperature. This is equivalent to considering only
the vibrational ``component'' of the H$_2$-H cooling rate. This approximation is valid only at
high temperatures: above $\sim$2000 K, the vibrational cooling function and the 
H$_2$ cooling function by \cite{glover2008} differ for at most a factor 3.5. The comparison
with the fit provided by
\cite{glover2008} can be significant only during these first phases of
the post-shock, when the low-density limit is satisfied and the gas temperature is high enough
that the vibrational transitions are expected to play a more relevant role than
the rotational ones. However, the assumed LTE population for the rotational transitions is
questionable at these density regimes. For this
reason, in order to verify the sensitivity of the vibrational
cooling function on several
assumptions on the rotational distribution, we have run the simulation with different hypothesis: ground
state approximation (where only the fundamental rotational level j=0 of each vibrational 
level $v$ is considered) and LTE rotational distribution with temperature smaller than
the effective gas temperature, in order 
to simulate the expected subthermal distribution at low densities. The results are shown in Fig.~\ref{figcomp} and they 
suggest that, as a first approximation, the effects on the vibrational cooling function
(and eventually on the chemistry) are negligible.

In order to model the non-equilibrium cooling it is useful to
compute the excitation temperature of the lowest transition
\be
T_{0-1}\equiv \frac{E_1-E_0}{k_{\rm B}\ln(n_{{\rm H}_2,0}/n_{{\rm H}_2,1})},
\label{excitationtemps}
\ee
where $E_1$ and $E_0$ are the energies of the first excited and
ground vibrational level and $n_{{\rm H}_2,1}$ and $n_{{\rm H}_2,0}$ their
abundances. Initially, the excitation temperature $T_{0-1}$ diverges
since the H$_2$ is still forming. In fact, according to the definition given in 
Eq.~\ref{excitationtemps} fractional abundances are very small and eventually the
logarithm diverges. For this reason, in Fig.~\ref{fig:temp} the excitation temperature is shown
starting from a fixed value of $t/t_H$ when the fractional abundances start to be 
significant. Once H$_2$
starts to form, the excitation temperature reaches a value higher
than the radiation and gas temperatures as shown in Fig.~\ref{fig:temp}.
Although the gas temperature changes very rapidly and varies over
several orders of magnitude, the excitation temperature T$_{0-1}$
tends rapidly to the final value of few hundreds K. Hence, while the gas and the radiation reach
thermal equilibrium, molecular vibration is decoupled as a result of the non-equilibrium patterns
in the vibrational distribution.

The state-to-state formation and
destruction pathways have been explicitely implemented in the kinetics. This
means that the energy deposited in the gas by exothermal chemical
processes is explicitly
included according to the kinetics itself. However,
this energy is not transferred back to the gas
and transformed into kinetic energy since the densities at which
such heating mechanisms are relevant are far from the present
conditions that are limited to $n \leq 10^8$~cm$^{-3}$.

\begin{figure*}
\begin{tabular*}{\textwidth}{c c}            
  \includegraphics[width=6cm,angle=-90]{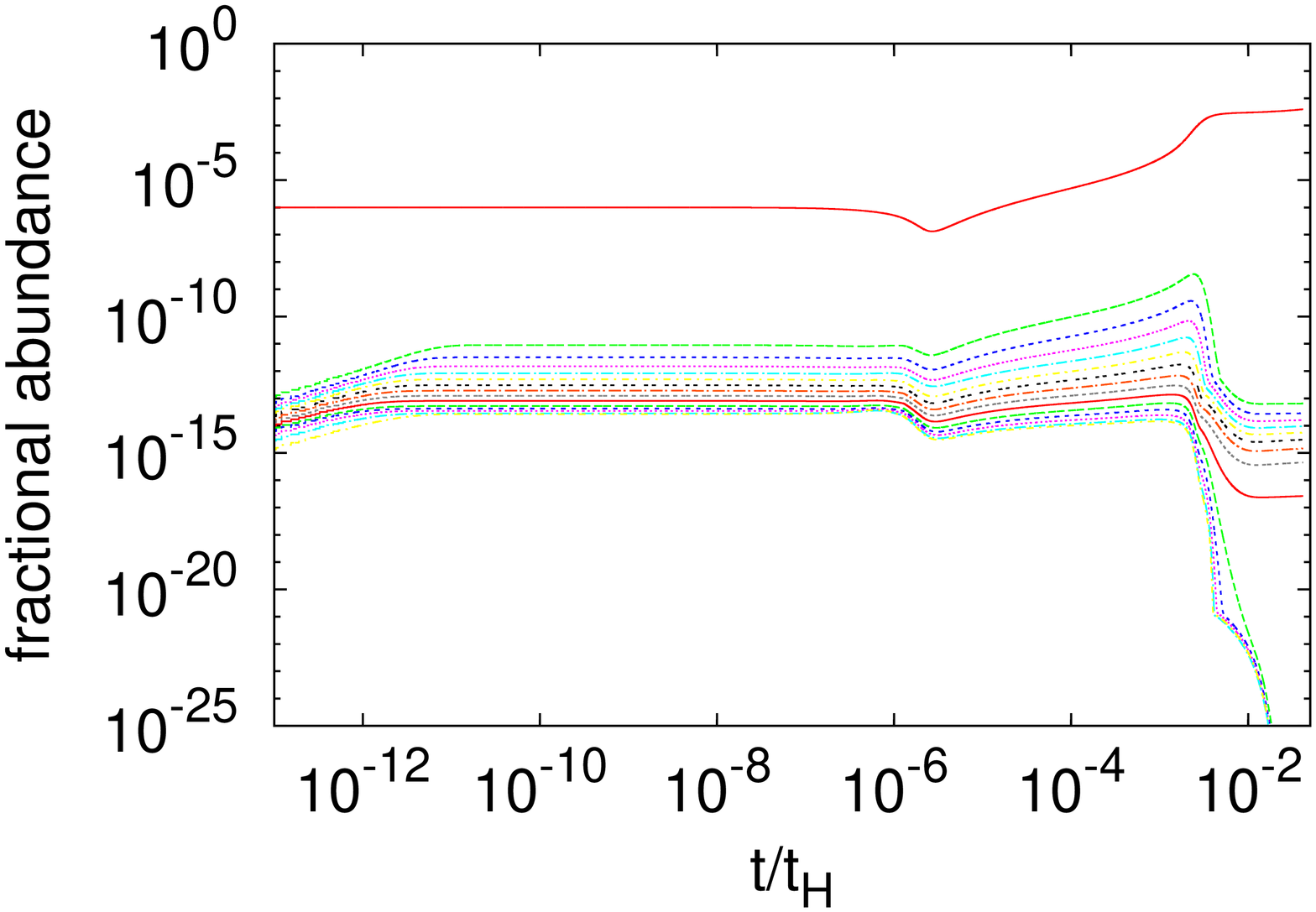} & \includegraphics[width=6cm,angle=-90]{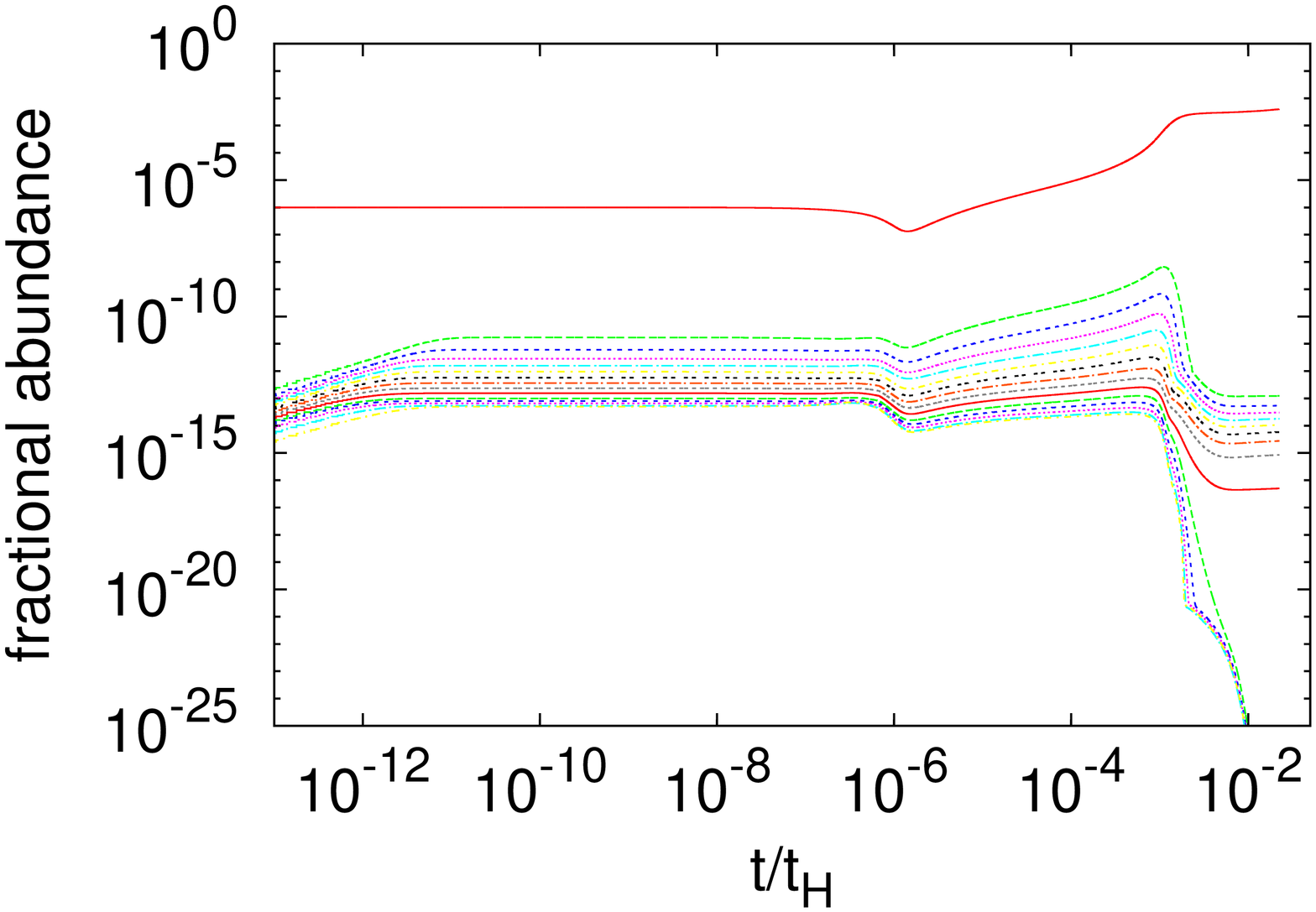}\\ 
  \includegraphics[width=6cm,angle=-90]{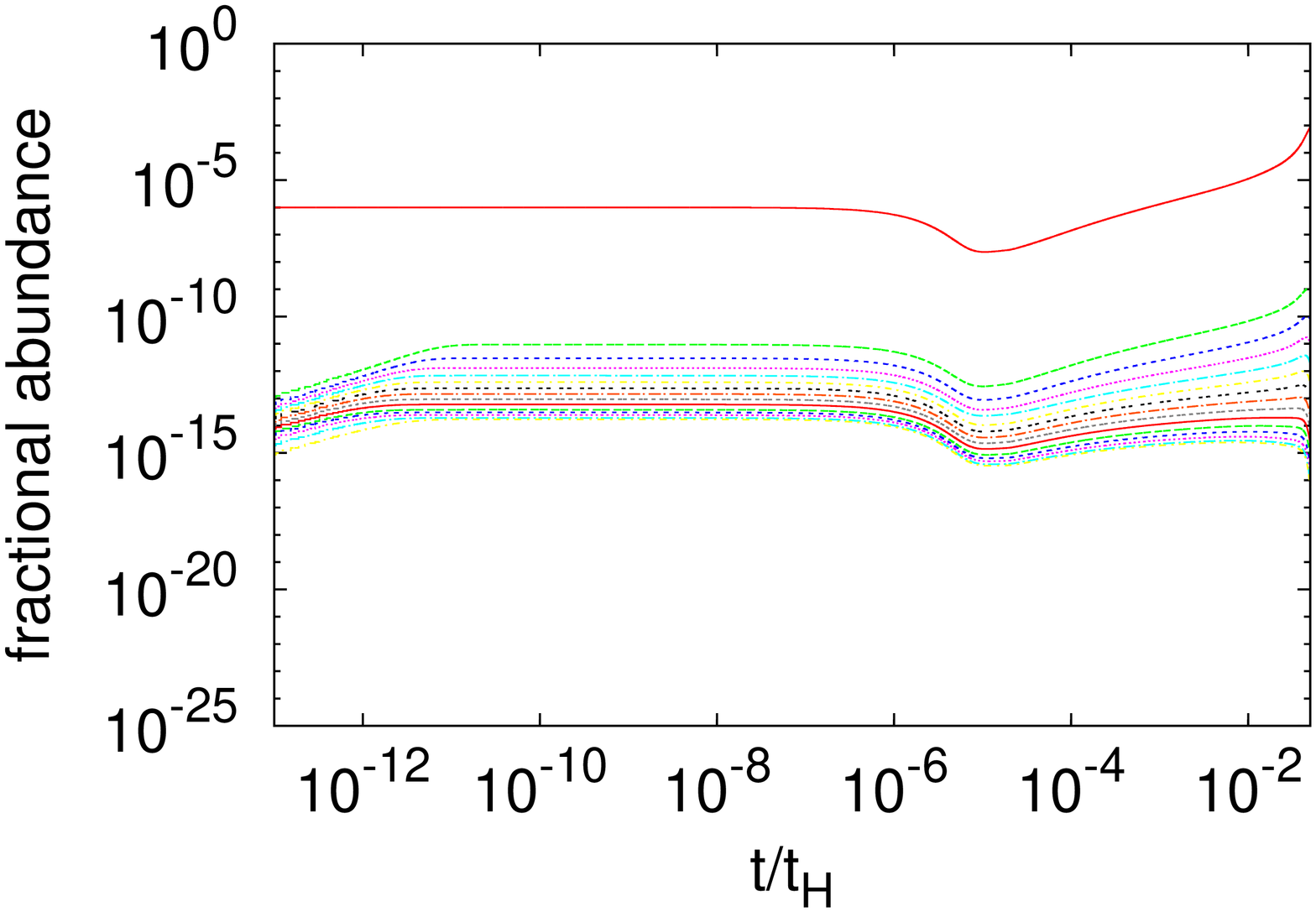} & \includegraphics[width=6cm,angle=-90]{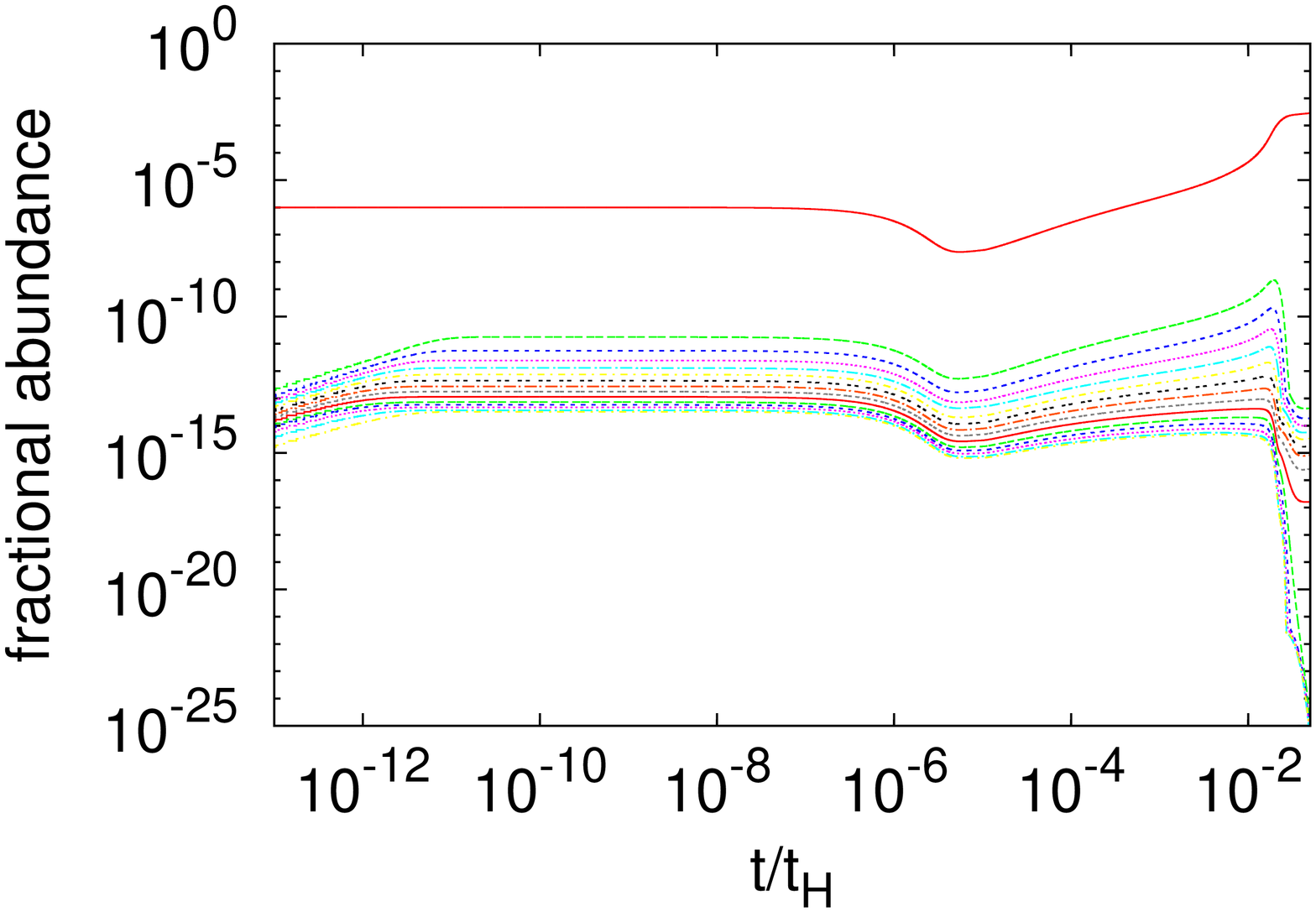}\\
\end{tabular*}
\caption{$\mathrm{H_2}$ vibrational level population for different initial condition for redshift $z$ and shock velocity $v_s$.
{\it Top panels, from left to right}: standard case ($z=20$, $v_s$=50 km/s), 
case 1 ($z=25$, $v_s$=50 km/s). {\it Bottom panels, from left to right}: case 2 ($z=20$, $v_s$=30 km/s), case 3 ($z=25$, $v_s$=30 km/s).
It can be noticed that a change in the redshift $z$ results in a time shit of the fractional abundance as 
a function of time; on the other hand, a change in the shock velocity affects the typical recombination scales.
For this reason in the case of higher shock velocities recombination proceeds slightly faster allowing to reach the 
freeze-out fractional abundances earlier in time.}
\label{comparison}
\end{figure*}

In order to provide an estimate for the sensitivity of our results
to redshift and shock velocity, we have run several models
with different initial conditions. The results are shown in Fig.~\ref{comparison}
and can be summarized as follows:
a change in the redshift $z$ 
implies a variation of the initial density that eventually produces a shift in time 
in the fractional abundances, leaving the characteristic timescales unchanged.
On the other hand,  a variation on the 
velocity of the shock affects the time scales of processes, pushing the recombination to be completed
at earlier times for faster shocks. The cases for $z=20, 25$ and $v_s=30-50$ km~s$^{-1}$
are shown in Fig.~\ref{comparison}.

\newpage

\section{Conclusions}

We have performed a numerical calculation of the chemical kinetics
for a shock-heated gas of primordial composition.
Particular attention has been paid to the distribution of the
vibrational levels of molecular hydrogen and its cation. Unlike
previous studies that assumed vibrational equilibrium for the level
population,  here we show that this assumption is not
always justified. In fact, in our calculations the vibrational level
populations are shown to deviate strongly from the equilibrium
predictions.
The case study considered here is that of a shock wave produced
by a primordial Supernova explosion. The results show that
H$_2$ cooling is crucial for lowering the temperature down
to $\sim 200$ K when the populations of the vibrational
levels of H$_2$ and H$^+_2$ molecules are out-of-equilibrium. As for
H$_2$, we find that the vibrational
distribution becomes unimportant as long as the temperature exceeds
$\sim10^3$~K. The first excited vibrational level has been found
to be at the equilibrium value 
for low temperatures,
but since the H$_2$ formation
tends to populate the higher vibrational levels, a non-equilibrium
distribution is produced for $v \leq 2$.

We plan to extend the present calculations changing the
chemical content of the gas allowing for the presence of heavy
elements. A rotationally resolved chemical kinetics should also be
considered allowing to relax the assumption of rotational equilibrium.
In this way we could be able
to properly describe any possible non-equilibrium effects on
the cooling function, related to the inclusion
of a state-to-state approach in the description of molecular systems, 
especially at low temperature.
Finally, optical depth effects may be included for a complete
treatment of the post-shock gas in a variety of environments.

\section*{Acknowledgments}
The Authors are grateful to the anonymous referee for the valuable comments
suggested. CMC, GM and SL acknowledge financial support of MIUR-PRIN (grant
no. 2010ERFKXL). This work has also been partially supported by the
FP7 project ''Phys4Entry'' - grant agreement n. 242311. CMC is
grateful to the CNR-IMIP computing facilities, in particular to Dr.
Pierpaolo Minelli for technical support. CMC, DG, SL and FP acknowledge
the discussions within the international team \#272 lead by C. M.
Coppola ``EUROPA - Early Universe: Research on Plasma Astrochemistry''
at ISSI (International Space Science Institute) in Bern.
GM acknowledges partial support from the EPSRC under grant number EP/L015374/1.

\appendix

\onecolumn

\section{Tables}
A list of the processes considered in the present
work can be found in the following Tables, together with analytical fits of 
the rate coefficients and the
references from which the data have been derived. For clarity,
we have defined $T_0=11604.5$~K, the temperature corresponding to 1~eV, 
and denoted the gas temperature 
and the radiation temperature (both in K) 
as $T$ and  $T_{\rm r}$, respectively. ln and log$_10$ represent the base $e$ and base $10$ logarithmics.

\begin{longtable}{|l|l|l|}
\caption{CHEMICAL REACTIONS AND RATE COEFFICIENTS.}\\
\hline
Chemical process & Rate coefficient (m$^3$~s$^{-1}$ or s$^{-1}$) & Ref. \\
\hline
\endfirsthead
\multicolumn{2}{r}%
{\tablename\ \thetable\ -- \textit{Continued from previous page}} \\
\hline
Chemical process & Rate coefficient (m$^3$~s$^{-1}$ or s$^{-1}$) & Ref. \\
\hline
\endhead
\hline \multicolumn{2}{r}{\textit{Continued on next page}} \\
\endfoot
\hline
\endlastfoot
1) $\mathrm{H}+\mathrm{e}^-\rightarrow \mathrm{H}^-+h\nu$				&$1.4\times10^{-24}T^{0.928}\exp(T/16200)$				& \cite{galli_chemistry_1998}\\
2) $\mathrm{H}^-+\mathrm{e}^-\rightarrow \mathrm{H}+2\mathrm{e}^-$	                &$1.27\times10^{-17}T^{1/2}\exp{(-157809.1/T)}(1+T_5^{1/2})^{-1}$ &\cite{black1981}\\
											&	                                         &\cite{haiman_h_1996}\\
3) $\mathrm{H}^-+\mathrm{H}\rightarrow 2\mathrm{H}+\mathrm{e}^-$	&for T $>$1160.45                         &      \cite{abel_modeling_1997}\\                                 
                                                                        &$10^{-6}\exp{[-20.37260896+1.13944933\ln{(T/T_0)}}+$&                   \\
						                        &$-0.14210135\ln{(T/T_0)}^2+$						&\\
									&$+8.4644554\times10^{-3}\ln{(T/T_0)}^3+$			&\\
									&$-1.4327641\times10^{-3}\ln{(T/T_0)}^4+$			&\\
									&$+2.0122503\times10^{-4}\ln{(T/T_0)}^5+$			&\\
									&$+8.6639632\times10^{-5}\ln{(T/T_0)}^6+$			&\\
									&$-2.5850097\times10^{-5}\ln{(T/T_0)}^7+$			&\\
									&$+2.4555012\times10^{-6}\ln{(T/T_0)}^8+$			&\\
									&$-8.0683825\times10^{-8}\ln{(T/T_0)}^9]$			&\\
									&for T $<$1160.45: 2.5634$\times10^{-15}\times$ T$^{1.78186}$                    &      \\
4) $\mathrm{H}^-+\mathrm{H}^+\rightarrow2\mathrm{H}$					&$1.4\times10^{-13}(T/300)^{-0.487}\exp(T/29300)$		&\cite{schleicher_2008}\\
5) $\mathrm{H}^-+h\nu\rightarrow \mathrm{H}+\mathrm{e}^-$				&$0.01\,T_{\rm r}^{2.13}\exp(-8823/T_{\rm r})$	& \cite{c11}\\
6) $\mathrm{D}^-+h\nu\rightarrow \mathrm{D}+\mathrm{e}^-$	& as (5) & \cite{c11}\\
7) HD$^+ + h\nu \rightarrow$  D  $+$    H$^+$ & $0.5\times 1.63\times 10^7\exp(-32400/T_{\rm r})$ & \cite{galli_chemistry_1998}\\
8) HD$^+ + h\nu \rightarrow$  H  $+$    D$^+$ & $0.5\times 1.63\times 10^7\exp(-32400/T_{\rm r})$ & \cite{galli_chemistry_1998}\\
9) HD$^+ + h\nu \rightarrow$  H  $^++$  D$^++$ e & 9$\times 10^1 T_{\rm r}^{1.48}\exp(-335000/T_{\rm r})$ & \cite{galli_chemistry_1998}\\
10) $\mathrm{H}+\mathrm{e}^-\rightarrow \mathrm{H}^++2\mathrm{e}^-$		&$10^{-6}\exp{[-32.71396786}+13.536556\ln{(T/T_0)}+$&\cite{abel_modeling_1997}\\
																		&$-5.73932875\ln{(T/T_0)}^2+$&\\
																		&$+1.56315498\ln{(T/T_0)}^3+$	&\\
																		&$-0.2877056\ln{(T/T_0)}^4+$&\\
																		&$+3.48255977\times10^{-2}\ln{(T/T_0)}^5+$	&\\
																		&$-2.63197617\times10^{-3}\ln{(T/T_0)}^6+$			&\\
																		&$+1.11954395\times10^{-4}\ln{(T/T_0)}^7+$			&\\
																		&$-2.03914985\times10^{-2}\ln{(T/T_0)}^8]$			&\\
11) H$^++e^-\rightarrow $H$+h\nu$ 											&2.753$\times$10$^{-20}$ (315614/$T$)$^{1.5}$(1+(115188/$T$)$^{0.407}$)$^{-2.242}$ &\cite{glover2009}\\
12) He$^{++}+$e$^-\rightarrow $He$^{+}+h\nu$ 									        &5.506$\times$10$^{-20}$(1262456/$T$)$^{1.5}$(1+(460752/$T$)$^{0.407}$)$^{-2.242}$ &\cite{glover2008}\\
13) He$^++$e$^-\rightarrow $He$+h\nu$ 	&10$^{-17}T^{-0.5}\cdot$                                                       &\cite{glover2008}\\
                                        &$(11.19-1.676 ($log$_{10}T)-0.2852($log$_{10}T)^2+0.04433($log$_{10}T)^3$)+    &  \\
                                        &+(1.9$\times$10$^{-9}T^{-1.5}\exp(-473421/T)$)(1+0.3$\exp(-94684/T)$) &  \\
14) HD + $h\nu \rightarrow$ HD$^+$ + e$^-$ & $2.9\times10^{2}T_{\rm r}^{1.56}\exp(-178500/T_{\rm r})$& \cite{galli_chemistry_1998}\\
15) D$^+$ + e$^-\rightarrow$ D+ $h\nu$ & as (11) &\\
16) D$^+$ + H $\rightarrow$ D + H$^+$ & $2.06\times 10^{-16}T^{0.396}\exp(-33/T)+2.03\times 10^{-15}T^{-0.332}$ & \cite{savin_2002}\\
17) H$^+$ + D $\rightarrow$ H + D$^+$ & $2 \times 10^{-16}T^{0.402}\exp(-37.1/T)-3.31\times 10^{-23}T^{1.48}$ & \cite{savin_2002}\\
18) D + H$\rightarrow$ HD + $h\nu$ & $10^{-32}[2.259-0.6(T/1000)^{0.5}+0.101(T/1000)^{-1.5}$&\\
&~~~~~~~~$-0.01535(T/1000)^{-2}+5.3\times 10^{-5}(T/1000)^{-3}]$&\cite{dickinson_2008}\\
19) HD$^+$ + H  $\rightarrow$ HD + H$^+$ & 6.4$\times 10^{-16}$ & \cite{stancil_1998}\\
20) D   + H$^+ \rightarrow$ HD$^+$ + $h\nu$ &$10^{-6}{\rm dex}[-19.38-1.523\log_{10}T+$ & \cite{galli_chemistry_1998}\\
& $+1.118(\log_{10}T)^2-0.1269(\log_{10}T)^3]$ & \\
21) H   + D$^+ \rightarrow$ HD$^+$ + $h\nu$ & as (19) &\\
22) HD$^+$ + e$^- \rightarrow$ D   + H & $7.2\times 10^{-14}T^{-0.5}$& \cite{stromholm_1995}\\
23) D  +  e$^-  \rightarrow$ D$^-$  + $h\nu$ & $3\times 10^{-22}(T/300)^{0.95} \exp(-T/9320)$& \cite{stancil_1998}\\
24) D$^+$ +  D$^- \rightarrow$ 2D &      $1.96\times 10^{-13}(T/300)^{-0.487}\exp(T/29300)$&\cite{lepp_2002}\\
25) H$^+$ +  D$^- \rightarrow$ D   + H & $1.61\times 10^{-13}(T/300)^{-0.487} \exp(T/29300)$&\cite{lepp_2002}\\
26) H$^-$ +  D  $\rightarrow$ H   + D$^-$ & $6.4\times 10^{-15}(T/300)^{0.41}$& \cite{stancil_1998}\\
27) D$^-$ +  H  $\rightarrow$ D   + H$^-$ & as (25) & \\
28) D$^-$ +  H  $\rightarrow$ HD  + e$^-$ & $1.5\times 10^{-21}(T/300)^{-0.1}$& \cite{stancil_1998} \\
29) D  +  H$^- \rightarrow$ HD  + e$^-$ &  as (23)& \cite{schleicher_2008} \\
30) H$^-$ +  D$^+ \rightarrow$ D   + H &$1.61\times 10^{-13}(T/300)^{-0.487}\exp(T/29300)$ & \cite{lepp_2002}\\
31) D    + H$_2 \rightarrow$ HD   + H  &  $T < 250$~K: $1.69 \times 10^{-16}\exp(-4680/T+198800/T^2)$& \cite{galli_2002}\\
                                             &  $T > 250$~K: $9\times 10^{-17}\exp(-3876/T)$& \cite{galli_chemistry_1998}\\
32) D$^+$   + H$_2 \rightarrow$ HD   + H$^+$ & $10^{-15}(0.417+0.846\log_{10}-0.137\log_{10}^2)$& \cite{galli_2002}\\
33) HD   + H  $\rightarrow$ D    + H$_2$ &  $T < 200$~K: $5.25\times 10^{-17}\exp(-4430/T+173900/T^2)$& \cite{galli_2002}\\
                                              &  $T > 200$~K: $3.2\times 10^{-17}\exp(-3624/T)$& \cite{galli_chemistry_1998}\\
34) HD   + H$^+ \rightarrow$ D$^+$   + H$_2$ & $1.1\times 10^{-15}\exp(-488/T)$& \cite{galli_2002}\\
35) He   + H$^+\rightarrow$ He$^+$    + H &  $T > 10^4$~K: $4\times 10^{-43}T^{4.74}$& \cite{galli_chemistry_1998}\\
                                              &  $T < 10^4$~K: $1.26\times 10^{-15} T^{-0.75}\exp(-127500/T)$& \cite{glover_2007}\\
36) H   + He$^+\rightarrow$ H$^+$    + He & $1.25\times 10^{-21}(T/300)^{0.25}$ & \cite{zygelman_1989}\\
37) He   +   H$^+\rightarrow$    HeH$^++h\nu$ & $8\times 10^{-26}(T/300)^{-0.24}\exp(-T/4000)$ & \cite{stancil_1998}\\
38) He   +   H$^++h\nu\rightarrow$    HeH$^++h\nu$ & $3.2\times10^{-26}[T^{1.8}/(1+0.1T^{2.04})]\times $ & \cite{jurek_1995}\\
& $\times \exp(-T/4000)(1+2\times 10^{-4}T_{\rm r}^{1.1})$ & \\
& &\cite{zygelman_1998} \\
39) He + H$_2^+\rightarrow$    HeH$^++$  H & $3\times 10^{-16}\exp(-6717/T)$& \cite{galli_chemistry_1998}\\
40) He$^+$ + H $\rightarrow$   HeH$^++h\nu$ & $4.16\times 10^{-22}T^{-0.37}\exp(-T/87600)$ & \cite{stancil_1998}\\
41) HeH$^+$+ H $\rightarrow$   He + H$_2^+$ & $0.69\times 10^{-15}(T/300)^{0.13}\exp(-T/33100)$ & \cite{linder1995}\\
42) HeH$^+$   + e  $\rightarrow$    He   +    H & $3\times 10^{-14}(T/300)^{-0.47}$ & \cite{stancil_1998}\\
43) HeH$^+ +h\nu \rightarrow$      He  +    H$^+$ & $220\times T_{\rm r}^{0.9}\exp(-22740/T_{\rm r})$ & \cite{jurek_1995}\\
44) HeH$^++h\nu \rightarrow$      H  +    He$^+$ & $7.8\times 10^3T_{\rm r}^{1.2}\exp^{-240000/T_{\rm r}}$ & \cite{galli_chemistry_1998}\\
45) H$_2$ + H$^+  \rightarrow$   H$_3^+ + h\nu$ & $10^{-22}$ & \cite{gerlich_1992}\\
46) H$_3^+$ +  e $\rightarrow$ H$_2$ + H & $0.34\times(1.27 \times 10^{-12}T^{-0.48}-1.3\times 10^{-14})$ & \cite{mccall_2004} \\
47) H$_3^+$ +  e $\rightarrow$ H + H + H & $0.66\times(1.27 \times 10^{-12}T^{-0.48}-1.3\times 10^{-14})$ & \cite{mccall_2004} \\
48) H$_2^+$ + H$_2\rightarrow$ H$_3^+$ + H & $2\times 10^{-15}$ & \cite{theard_1974}\\
49) H$_3^+$ + H  $\rightarrow$ H$_2^+$ + H$_2$ & $7.7\times 10^{-15}\exp(-17560/T)$ & \cite{sidhu_1992}\\
50) H$_2$ + He$^+ \rightarrow$ He + H + H$^+$  & 2.7$\times 10^{-14}T^{-1.27}\exp(-43000/T)$ & \cite{glover2008} \\
51) H$_2$ + He$^+ \rightarrow$ H$_2^+$ + He    & 3.7$\times 10^{-20}\exp(35/T) $               & \cite{glover2008} \\
52) He + e   $\rightarrow$ He$^{+}$+ e + e   & 2.38$\times 10^{-17}T^{0.5} \exp(-285335.4/T)$ &\cite{black1981} \\
					          & $\times (1+\sqrt{T/10^5})^{-1}$                &	\cite{haiman_h_1996}			\\
53) He$^+$ + e       $\rightarrow$ He$^{++}$+ e + e  & 5.68$\times 10^{-18}T^{0.5} \exp(-631515/T)$ &\cite{black1981} \\					          
					          & $\times (1+\sqrt{T/10^5})^{-1}$                &	\cite{haiman_h_1996}			\\
\label{tab:atomic}
\end{longtable}

\clearpage

\begin{table*}
\centering
\begin{minipage}{160mm}
\begin{tabular}{ll}
\hline
	& Chemical process  \\
\hline
1	& $\mathrm{H}+\mathrm{H}_2(v)\rightarrow\mathrm{H}+\mathrm{H}_2(v')$\\
2	& $\mathrm{H}+\mathrm{H}^-\rightarrow\mathrm{H}_2+\mathrm{e}^-$\\
3	& $\mathrm{H}+\mathrm{H}^+\rightarrow\mathrm{H}_2^++h\nu$\\
4	& $\mathrm{H}_2^+(v)+\mathrm{H}\rightarrow\mathrm{H}_2(v')+\mathrm{H}^+$\\
5	& $\mathrm{H}_2(v)+\mathrm{H}^+\rightarrow\mathrm{H}_2^+(v')+\mathrm{H}$\\
6	& $\mathrm{H}+\mathrm{H}_2^+(v)\rightarrow\mathrm{H}+\mathrm{H}_2^+(v')$\\
7	& $\mathrm{H}^++\mathrm{H}_2(v)\rightarrow\mathrm{H}^++\mathrm{H}_2(v')$\\
8	& $\mathrm{H}+\mathrm{H}_2(v)\rightarrow\mathrm{H}+\mathrm{H}_2(v')$\\
9	& $\mathrm{H}^++\mathrm{H}_2(v)\rightarrow\mathrm{H}+\mathrm{H}+\mathrm{H}^+$\\
10	& $\mathrm{H}_2(v)+h\nu\rightarrow\mathrm{H}_2^+(v')+\mathrm{e}^-$\\
11	& $\mathrm{H}_2^+(v)+h\nu\rightarrow\mathrm{H}+\mathrm{H}^+$\\
12	& $\mathrm{H}_2(v)+h\nu\rightarrow\mathrm{H}+\mathrm{H}$\\
13	& $\mathrm{H}_2^+(v)+\mathrm{e}^-\rightarrow\mathrm{H}+\mathrm{H}$\\
14	& $\mathrm{H}_2(v)+\mathrm{e}^-\rightarrow\mathrm{H}^-+\mathrm{H}$\\
15	& $\mathrm{H}_2(v)+\mathrm{e}^-\rightarrow\mathrm{H}_2^*\rightarrow\mathrm{H}_2(v')+\mathrm{e}^-+h\nu$\\
16	& $\mathrm{H}_2(v)\rightarrow\mathrm{H}_2(v')+h\nu$\\
17	& $\mathrm{H}_2^+(v)\rightarrow\mathrm{H}_2^+(v')+h\nu$\\
18	& $\mathrm{H}+\mathrm{H}_2(v)\rightarrow\mathrm{H}+\mathrm{H}+\mathrm{H}$\\
\hline
\end{tabular}
\end{minipage}
\caption{Molecular processes: vibrationally resolved reactions. All data for these processes are from \protect\cite{c11}.}
\label{tab:molecular}
\end{table*}

\clearpage

{\captionsetup[table]{aboveskip=2cm}
\begin{sidewaystable*}
\begin{minipage}{160mm}
\begin{tabular}{llll}
\hline	
&Process & J~m$^3$~s$^{-1}$ or J~s$^{-1}$& Reference\\
\hline
{\bf Atomic processes}                  &  & &\\
&{\bf -collisional excitation} &&\\
&~H                               & $7.50\times10^{-32}\exp{(-118348/T)}(1+T_5^{1/2})^{-1}$                 & \protect\cite{anninos_cosmological_1997}\\
&{\bf -collisional ionisation} &&\\
&~H                               & $1.27\times10^{-34}T^{1/2}\exp{(-157809.1/T)}(1+T_5^{1/2})^{-1}$            & \protect\cite{haiman_h_1996}\\
&~He                              & $9.38\times10^{-35}T^{1/2}\exp{(-285335.4/T)}(1+T_5^{1/2})^{-1}$            & \protect\cite{haiman_h_1996}\\
&~He                              & $4.95\times10^{-35}T^{1/2}\exp(-631515./T)(1+T_5^{1/2})^{-1}$         & \protect\cite{haiman_h_1996}\\
&{\bf -bremsstrahlung}            & $1.42\times10^{-40}T^{1/2}[1.10+0.34\exp{(-(5.50-\log_{10}{T})^2/3)}]$  & \protect\cite{haiman_h_1996}\\
&{\bf -Compton cooling/heating}   & $1.017\times10^{-44}T_{\rm r}^4(T-T_{\rm r})$                           & \protect\cite{anninos_cosmological_1997}\\
&{\bf -recombination} &&\\
&~H$^+$                           & $8.70\times10^{-40}T^{1/2}T_3^{-0.2}[(1+T_6^{0.7})^{-1}]$           & \protect\cite{haiman_h_1996}\\
&~He$^+$                          & $1.55\times10^{-39}T^{0.3647}$                                     & \\
&                                 & $+1.24\times10^{-26}T^{-1.5}(1+0.3\exp{(-94000/T)})\exp(-470000/T)$& \protect\cite{haiman_h_1996}\\
&~He$^{++}$                       & $3.48\times10^{-39}T^{1/2}T_3^{-0.2}(1+T_6^{0.7})^{-1}$         & \protect\cite{haiman_h_1996}\\
&{\bf -e$^-$ attachment to H}     & &  calculated using reaction rate 1)\\
{\bf Molecular processes}         &             &               &\\
&{\bf collisional cooling}                     &&\\
&~H$_2$/H                          & see reference            & \protect\cite{glover2008}\\
&~H$_2$/H$_2$                      &  //                      & \protect\cite{glover2008}\\
&~H$_2$/He                         &  //                      & \protect\cite{glover2008}\\
&~H$_2$/H$^+$                      &  //                      & \protect\cite{glover2015}\\
&~H$_2$/e$^-$                      &  //                      & \protect\cite{glover2015}\\
&{\bf HD cooling}                  & $\Lambda(T,n)$           & \protect\cite{lipovka_cooling_2005}\\
{\bf Heating process}         &             &               &\\
&H$^-$ channel for H$_2$ formation & vibrationally resolved (Eq.~\ref{gamma})&  cross-sections by \protect\cite{cizek_1998}\\
\hline
\end{tabular}
\end{minipage}
\caption{Cooling functions included in the thermal evolution. The symbol $T_n$ stands for $T/10^n$.}
\label{tab:cooling}
\end{sidewaystable*}

\bibliographystyle{mn2e}
\bibliography{biblio}

\end{document}